# Splitting of transmission peak due to the hole symmetry breaking

Xiao-gang Yin, Cheng-ping Huang, Zhi-Qiang Shen, Qian-jin Wang,

and Yong-yuan Zhu*

*National Laboratory of Solid State Microstructures, Nanjing University*

*Nanjing 210093, P.R. China*

Abstract

We studied experimentally and theoretically the optical transmission through asymmetrical holes of a metal film, which is constructed by introducing small protuberances to the sides of individual square holes. Due to the symmetry breaking of hole shape, an interesting transmission feature appears: both the Ag-glass (1, 0) and Ag-glass (1, 1) peaks split distinctly. Detailed studies indicate that the peak splitting is actually associated with the two asymmetrical waveguide surface-plasmon (WSP) modes confined on the surface of opposite hole walls. The finding demonstrates the crucial role of WSP modes and enriches our understanding of the phenomenon.

Email: yyzhu@nju.edu.cn

Recently, tremendous attention has been focused on the periodically perforated metal film which exhibits an extraordinary optical transmission (EOT) [1]. Besides the interest in exploring the potential applications of the effect, many studies have been devoted to the understanding of the underlying physics. Although there are several mechanisms proposed [2-4], it is generally acknowledged that the surface-plasmon polaritons (SPPs) modes originating from the coupling of light with the surface-charge oscillation play a dominant role [5-7]. This explanation is partly in accordance with the experimental and theoretical results that the position of transmission peak is sensitive to the period of hole lattice.

On the other hand, the study by Klein *et al*. [8] has demonstrated that the light transmission can be greatly influenced by the shape and size of the subwavelength holes. Similar effect has also been suggested for the single holes milled into a metal film [9, 10]. It was believed that the localized surface-plasmon (LSP) modes excited on the hole ridge are responsible for the effect [9, 11]. Nonetheless, the theoretical calculations have also highlighted the importance of waveguide modes, where the Fabry-Perot resonance of a single hole will present a transmission peak [12, 13]. Compared with a perfect waveguide, the waveguide mode in a real-metal case generally has a SPP character: the mode is laterally evanescent and confined on the surface of hole walls [4, 14, 15]. This SPP-like waveguide mode can be termed as the waveguide surface-plasmon (WSP) mode, which is mainly dependent on the shape and size of waveguide (note that the WSP mode is distinct from the aforementioned LSP mode). Certainly, the WSP mode will also play an important role in the light transmission through the hole arrays, where the energy is mediated along the hole depth either propagating or evanescently relying on the operating and the cutoff wavelength. Up to now, most of the previous studies have employed the metal films perforated with the symmetrical holes, such as the rectangular [8, 11], elliptical [16], equilateral triangular [17], square [18], and annular [19] holes etc. In these cases, the WSP or waveguide mode is usually excited with a symmetrical field distribution.

In this paper, we aim at the optical properties of asymmetrical air holes which are drilled in a square array in a metal film. Explicitly, the asymmetrical holes studied here are constructed by introducing small metallic protuberances to one or two sides of the individual square holes. Compared with the original square hole array, an interesting transmission feature appears: both the Ag-glass (1, 0) and Ag-glass (1, 1) peaks split distinctly. We found that the peak splitting is associated with the two asymmetrical WSP modes confined on the surface of asymmetrical hole walls.

Figure 1(a) shows one part of the designed structure with the structural parameters depicted in Fig. 1(b). In the experiment, a 70 nm-thick silver film deposited on the glass substrate is prepared (although the film is not completely opaque, previous study has showed that the light transmission is mainly mediated by the holes rather than the film itself [20]) and the designed patterns are fabricated using the focused-ion-beam (FIB) system (Strata FIB 201, FEI Company, 30 KeV Ga ions, 10 pA beam current). The typical FIB image of a part of the sample (each array consists of 66×66 units) is shown in Fig. 1(c). The zero-order transmission intensities, which will be normalized to that of the bare glass substrate, are collected by an optical spectrum analyzer (ANDO AQ-6315A). Here, the light is incident normally onto the sample with the electric field along the *x*- or *y*-axis (called *x*- or *y*-polarization, respectively). In addition, numerical simulations have been performed with the finite-difference time-domain (FDTD) method. In the simulations, the permittivity of glass substrate is set as 2.25, and the silver is modeled by the Drude dispersion with a plasma frequency of $\omega_p = 1.37 \times 10^{16}$ rad/s and a collision frequency of $\gamma = 2 \times 10^{14}$ rad/s. The latter is larger than that of the bulk metal, due to additional scattering from the nanoscale surface.

Without loss of generality, figure 2 shows, for the *y*-polarization, both the measured (a) and calculated (b) zero-order transmission spectra of a proposed structure (red lines) as well as of a referenced pure square hole array (without protuberance, black lines). Here the lattice period is *p*=600 nm and the hole side is *a*=300 nm. For the protuberance, the width is *w*=150 nm, and the length is $l_1$=0 and $l_2$=60 nm. Note that all the calculated transmission spectra agree well with the experimental results. The black lines in Figs. 2(a) and 2(b) exhibit a set of transmission peaks, which can be attributed to the SPP resonance. However, when one small protuberance is introduced to the right hole wall, an interesting transmission feature is observed (Fig. 2, red lines). The Ag-glass (1, 0) SPP peak splits into two transmission peaks, named as (1, 0)-1 peak and (1, 0)-2 peak, respectively (as marked by the arrows). Apparently, the (1, 0)-1 peak is very close to the position of Ag-glass (1, 0) SPP resonance, while the (1, 0)-2 peak is on the right-hand side of this position. Moreover, one can see that the Ag-glass (1, 1) SPP peak also splits into two peaks, named as (1, 1)-1 peak and (1, 1)-2 peak, respectively (as marked by the arrows). On the other hand, when the incident light is x-polarized, the transmission spectrum is almost identical to that of the pure square hole array and no peak splitting can be found (not shown here). Here a question rises: what is the origin of the splitting of transmission peaks?

To answer the question, we first check the fundamental waveguide mode of the original square hole without the protuberance. Following the result of Ref. [4], the dominant electric field in the waveguide (for the y-polarization) is found to be $E_y \propto A(e^{\alpha y} + e^{-\alpha y})$, where $|y| \leq a/2$ and $\alpha = k_0 \sqrt{2\delta/a}$ ($k_0$ is the wavevector in free space and $\delta$ is the skin depth of the metal). Thus, the waveguide modes are bounded to the hole walls ($y = \pm a/2$) due to the excitation of electronic oscillations [14, 15]. In other words, the waveguide mode is a superposition of two WSP modes with the symmetrical field amplitudes. In the light transmission, the two WSP modes can be equally excited by the surface waves (SPP mode). Nonetheless, when the protuberance is introduced to the square hole, the two WSP modes will be no longer symmetrical because of the symmetry breaking of the hole shape. In this case, the electric field can be expressed approximately as $E_y \propto A_1 e^{\alpha_1 y} + A_2 e^{-\alpha_2 y}$, where the amplitudes of WSP modes are determined according to the interplay between the surface waves and WSP modes themselves [4]. It should be mentioned that, compared with the WSP mode on the original hole wall, the WSP mode around the protuberance has a larger cutoff wavelength. This point has been suggested by the previous studies on the single C-shaped holes [21-23], where the transmission peak (close to the cutoff wavelength) is accompanied by an enhanced field localized near the protuberance.

Although the WSP mode around the protuberance may be propagating along the hole depth (due to an increase of cutoff wavelength [22, 23]), this character is actually not critical for the observed effect, as the metal film studied here is very thin and the phase retardation (or field decay) is not significant. Moreover, numerical simulations suggest that increase of film thickness will blueshift the split transmission peaks (not shown here), which is different from the waveguide resonance. To establish the possible relationship between the split peaks and the two WSP modes, we have plotted with the FDTD method the electric field patterns (Fig. 3) on the central plane $z$=-35nm, for the split peaks marked in Fig. 2. We stress that the obtained characteristic of the field distribution is maintained throughout the hole depth. One can see from Fig. 3 that the waveguide mode is generally composed of two evanescent waves localized on the left- and right-hand side, in agreement with the above analysis. However, the amplitude contrast between them is significant, giving rise to a minimum of field with the position deviating from the center of waveguide. Figure 3(a) shows that the $E$-field of (1, 0)-1 peak is mainly concentrated on the hole's left side, suggesting a coupling between the Ag-glass (1, 0) SPP mode and the left-side WSP mode. Instead, Figs. 3(b) shows that the $E$-field of (1, 0)-2 peak is mainly confined on the protuberance, indicating a coupling of Ag-glass (1, 0) SPP mode with the right-side WSP mode. Similar results can also be obtained with Fig. 3(c) and 3(d), where the Ag-glass (1, 1) SPP mode couples respectively to the left- and right-side WSP modes. Thus, we conclude that the peak splitting stems from the two asymmetrical WSP modes: the coupling between one WSP mode and one SPP mode will lead to an enhanced field as well as a transmission peak.

Figures 4(a) and 4(b) present, respectively, the measured and calculated spectra for the structure with different length $l_2$ of the right protuberance (the incident light is y-polarized and the left protuberance length is $l_1$=0). The theory and experiment agree well with each other. As $l_2$ increases from 20nm to 40nm and 60nm, the peak of (1, 0)-1 does not shift, but the peak of (1, 0)-2 shifts drastically to the longer wavelength. The same is with the (1, 1) peak. This can be understood, as the cutoff wavelength of right-side WSP mode is very sensitive to the protuberance length [22]. In addition, Figs. 4(c) and 4(d) present, respectively, the measured and calculated spectra for the structure with different length $l_1$ but $l_2$ is fixed as 60nm (for the $y$-polarization). As $l_1$ increases, the (1, 0)-1 and (1, 1)-1 peaks redshift gradually, as expected. Moreover, the (1, 0)-2 and (1, 1)-2 peaks also redshift, as the right-side WSP mode is increasey perturbated by the left protuberance. More interestingly, when the length of the left protuberance grows to 60 nm, the same as the right one, the four split transmission peaks will be merged into two peaks. That is, the peak splitting will not be present. In this case, the hole becomes a symmetrical I- or H-shaped hole [24], with the two WSP modes simultaneously and equally excited around the two protuberances.

The results above vividly indicate that the peak splitting is associated with the symmetry breaking of the hole shape. When the hole is symmetric, the two (or more) WSP modes confined to the opposite sides of the individual holes are degenerate, and no transmission peak splits. When the opposite sides of the individual holes are different, the two WSP modes will be asymmetrical and their interactions with the surface waves are also distinct in, such as the field distribution, the coupling strength, and the spectral position of field enhancement. This leads to a splitting of transmission peak, i.e., the generation of two transmission peaks with one SPP mode. The conclusion is also true to the protuberances of nonmetallic materials. For example, if the metallic protuberances are replaced by the dielectric ones (e.g., $SiO_2$), the peak splitting also presents (not shown here).

In conclusion, we have experimentally and theoretically investigated the splitting of transmission peaks with a square array of asymmetrical air holes. The peak splitting is attributed to the asymmetrical WSP modes due to the symmetry breaking of the hole shape. The finding confirms the important role of WSP modes in the optical transmission and will further people's understanding of the EOT phenomenon. Moreover, by employing the asymmetrical air holes, new optical devices supporting the multiple-wavelength output may be designed and constructed.

This work was supported by the State Key Program for Basic Research of China (Grant Nos. 2004CB619003 and 2006CB921804), by the National Natural Science Foundation of China (Grant No. 10523001, 10874079 and 10804051).

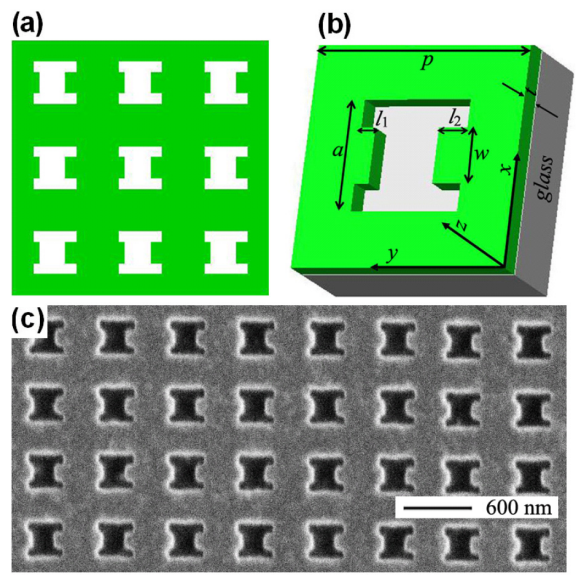

FIG.1. (Color online) (a) Schematic view of a perforated metallic film with nine unit cells. (b) The unit cell with the corresponding structural parameters, and (c) The focused ion beam (FIB) image of a typical array of the structure (the lattice period is $p$=600 nm). All samples are with a thickness $t$=70 nm.

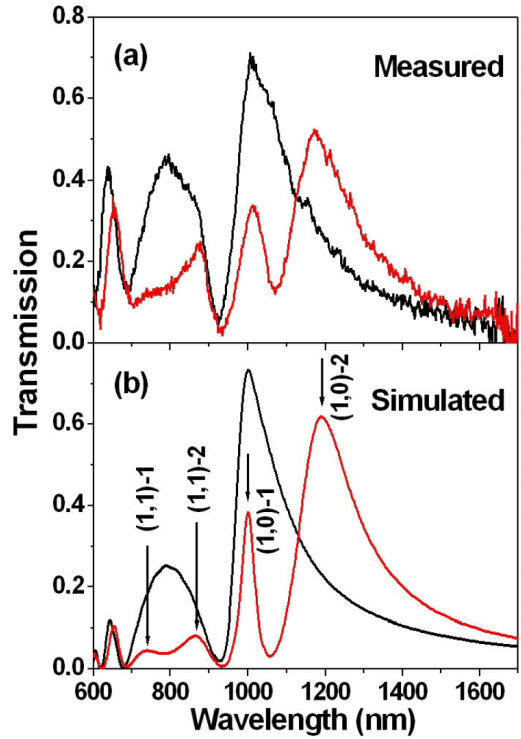

FIG.2. Measured (a) and calculated (b) zero-order transmission spectra, for $y$-polarization. The red lines correspond to a sample with the following parameters: $p$=600 nm, $a$=300 nm, $w$=150 nm, $l_1$=0 nm, and $l_2$=60 nm. And the black lines correspond to the referenced pure hole array without protuberance (with the same lattice period and hole size). The split transmission peaks are marked with the arrows.

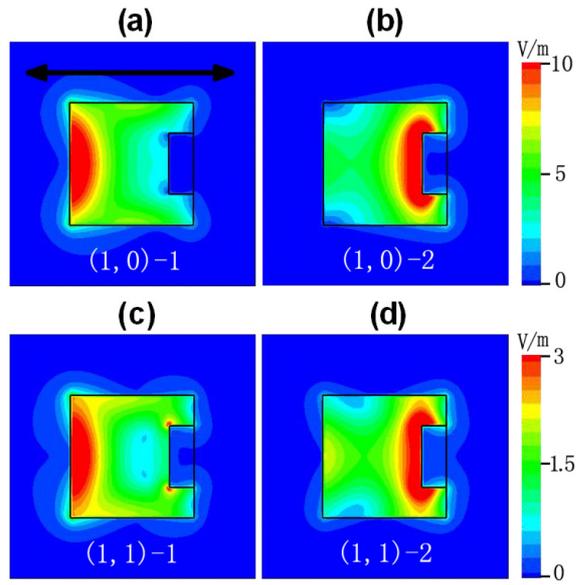

FIG.3. (Color online) Simulated electric field pattern on *x-y* plane ($z$=-35 nm), for the split transmission peaks (marked by the arrows in Fig. 2(b)). The bold solid arrow specifies the incident polarization (*y*-polarization). The image covers a single period, and the black lines mark the profiles of the square holes and protuberances.

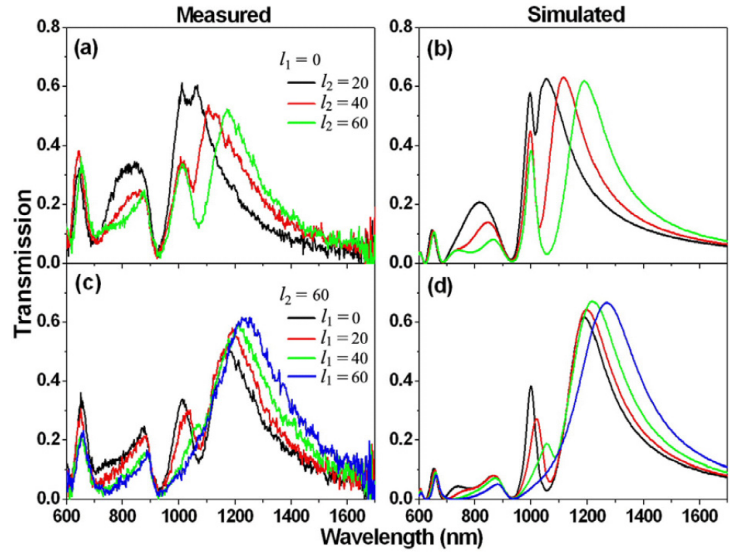

FIG.4. Measured (left column) and calculated (right column) zero-order transmission spectra, for *y*-polarization. In the first row, the length, $l_2$, of the right protuberances is changed from 20 to 60 nm, and the remaining parameters are fixed as $p$=600 nm, $a$=300nm, $w$=150 nm, and $l_1$=0; In the second row, the length, $l_1$, of the left protuberances is changed from 0 to 60 nm, and the remaining parameters are fixed as $p$=600 nm, $a$=300nm, $w$=150 nm, and $l_2$=60.